\documentclass[sigchi]{acmart}

\usepackage{booktabs} 
\usepackage[subtle]{savetrees}

\setcopyright{acmcopyright}

\begin{document}
\title[Data is Personal]{Data is Personal: Attitudes and Perceptions of Data Visualization in Rural Pennsylvania}

\author{Evan M. Peck}
\affiliation{%
  \institution{Bucknell University}
}
\email{evan.peck@bucknell.edu}

\author{Sofia E. Ayuso}
\affiliation{%
  \institution{Bucknell University}
}
\email{sea018@bucknell.edu}

\author{Omar El-Etr}
\affiliation{%
  \institution{Bucknell University}
}
\email{omar.eletr@bucknell.edu}

\begin{abstract}
Many of the guidelines that inform how designers create data visualizations originate in studies that unintentionally exclude populations that are most likely to be among the ``data poor''. In this paper, we explore which factors may drive attention and trust in rural populations with diverse economic and educational backgrounds - a segment that is largely underrepresented in the data visualization literature. In 42 semi-structured interviews in rural Pennsylvania (USA), we find that a complex set of factors intermix to inform attitudes and perceptions about data visualization - including educational background, political affiliation, and personal experience. The data and materials for this research can be found at \url{https://osf.io/uxwts/}
\end{abstract}

%
%




\newcommand{\todo}[1]{\textcolor{red}{TODO: #1}}
\newcommand{\note}[1]{\textcolor{blue}{NOTE: #1}}
\newcommand{\tocite}[1]{\textcolor{green}{CITE: #1}}

\maketitle

\section{Introduction}
Access to data can provide insight into our political, social, and physical environment. Following the development of web-based creation tools~\cite{Bostock2011, satyanarayan2016reactive, satyanarayan2017vega, wongsuphasawat2016voyager}, the recent prevalence of data visualization on the web has ushered in new-found hope for broadly accessible, engaging visualizations that can empower everyday people to understand and reason with data~\cite{viegas2006communication}. But what are the implications for people that do not pay attention to or understand this data? 

\begin{figure}
\includegraphics[width=3in]{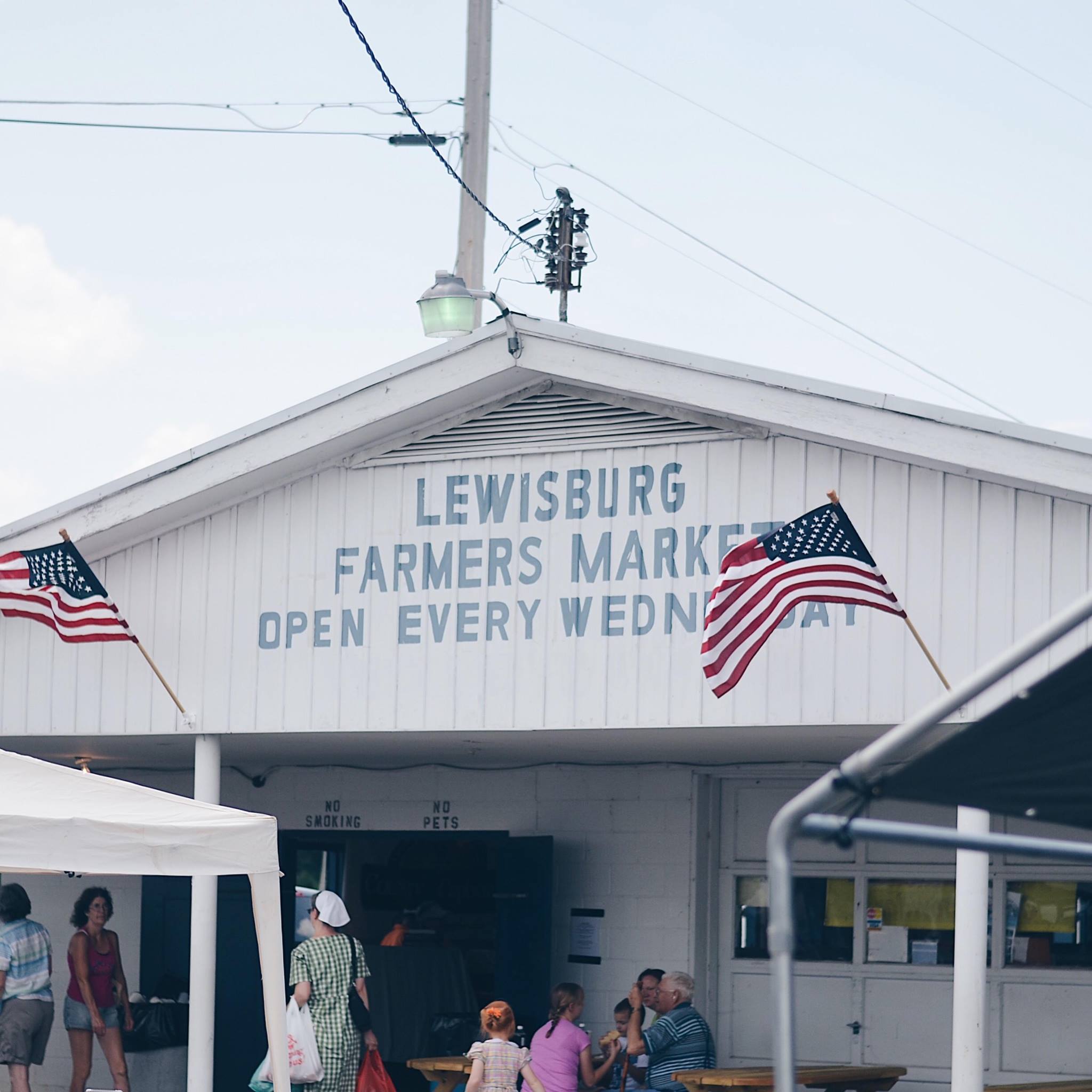}
\caption{We interviewed 42 community members in rural PA about their perceptions of data visualization. Above: Lewisburg Farmers market - one of our interview sites}
\end{figure}

Despite the emergence of data on the web, people often do not maximize these new opportunities for engagement~\cite{Boy2017, Koesten2017, Kong2018}, with potentially dire consequences~\cite{Burrell2018}. Yu contends that the ``information poor'' are disadvantaged in many dimensions that can fundamentally alter their engagement with society - lack of economic resources, basic literacy and skills, necessary service infrastructure, among others~\cite{Yu2010,Yu2011}. 

Rural people are particularly vulnerable to information paucity due to economic and infrastructure challenges~\cite{Burrell2018}. In comparison to their urban or suburban counterparts, rural populations tend to see gaps in education, income, device availability, and internet access~\cite{Hamby2018, Perrin2017, Rainie2004}. For example, Burrell remarks that digital inequality in rural regions is fundamentally a matter of exclusion~\cite{Burrell2018}.

Initiatives to broaden the accessibility of visualization are hardly novel. However, while traditional efforts to improve information and visualization literacy typically focus on education initiatives~\cite{Alper2017, gilbert2005visualization, gilbert2008visualization}, rural education faces  infrastructure and funding challenges that make large-scale changes unlikely for the near future. As information visualization continues to serve as a mediator for everyday people to understand how data describes and dictates their lives, it becomes important to question whether findings formed in laboratory settings still apply to audiences in hard-to-access communities that with diverse economic and educational backgrounds.  

In this paper, we share the results of 42 interviews with residents of rural Pennsylvania about their attitudes towards data visualization in an effort to begin articulating the factors that drive both perceptions of data, as well as attention towards data visualizations.

\section{Background}
Communicating data to audiences with diverse backgrounds comes with a variety of challenges. Encounters with data can be manipulated by several factors, including experience or education~\cite{Elias2011,Grammel2010,Grammel2012,Lee2016,Maltese2015,Parsons2017,Valdez2017}, biases~\cite{Bedek2017,Kong2018,Mansoor2017,Parsons2017,Valdez2018,Wall2017}, and attention~\cite{Dimara2017, Kennedy2017}. Kennedy and Hall consider that several different attitudes may mediate interaction with data - about subject matter, source or original location, and self-perceived skills to decode visualizations~\cite{Kennedy2017}. Our focus on people in rural settings is motivated by the population's absence in the visualization literature, and that gaps in education, income, and literacy~\cite{Burrell2018,Hamby2018} may impact perceptions of data visualizations. 

\subsection{Which visualizations do people understand?}
Accounting for a person's prior experience with data visualization is a challenge that can undermine even the most basic attempts to communicate data~\cite{Maltese2015, Lee2017, Boy2014, Lengler2006}. \textit{Visualization literacy} most often refers to the capability of a person ``to read, comprehend, and interpret'' graphs~\cite{Lee2017}. While education is often pointed to as a driver of poor visualization literacy, new graphic representations that are not accompanied by training can cause problems for people regardless of their background~\cite{Lee2016, Maltese2015, Grammel2010, Elias2011}. Most attempts at making data visualizations more broadly accessible focus on various components of education~\cite{gilbert2005visualization, gilbert2008visualization}. For example, Ruchikachorn and Mueller promote literacy through morphing one design into another ~\cite{Ruchikachorn2015} and Alper et al. created a tablet-based web application to teach data visualization to K-12 students~\cite{Alper2017}.

Lack of familiarity may also drive surprising results within the data visualization community. In one telling example, graphs were used to communicate information between doctors and patients. Hakone et al. found that pie charts that sampled temporal dimensions were more effective at communicating change-over-time data when compared to temporal area charts among older patients~\cite{Hakone2017}.

\subsection{Which visualizations do people pay attention to?}
While data visualization research typically investigates people who have already engaged with a graph, it is also critical to identify obstacles that impede interest or attention towards data in the first place. These issues arise more dramatically in the web's attention economy, where websites may have as little as 50 milliseconds to make a positive impression on users~\cite{Lindgaard2006}, and opinions about infographics may be formed within 500 ms~\cite{harrison2015infographic}. \textit{How can we drive attention to data without compromising the integrity of the data?}

In the context of data visualization, the answer to \textit{what drives attention} is multifaceted and difficult to summarize succinctly. For example, there have been renewed efforts in understanding the impact of emotion \textit{on} data visualization~\cite{Harrison2013}, as well as emotion \textit{in} data visualization~\cite{Kennedy2017, Boy2017}. These investigations often center around the value of visually rich representations of information (such as infographics), even if some forms distort or distract from the data. Pictographs and isotypes have studied within the context of helping communicate Bayesian reasoning~\cite{ottley2016improving}, eliciting empathy~\cite{Boy2017}, and improving engagement~\cite{Haroz2015}. Focusing on indices of recall and recognition, Borkin et al. investigates the topic of \textit{memorability} within the context of data visualizations~\cite{Borkin2013,Borkin2015}, following a study by Bateman et al.~\cite{Bateman2010}. The core takeaway is that graph titles and visual embellishments may facilitate long-term retention that extend beyond in-the-moment understanding.

A balance is needed, however, to avoid skewing data, as biases that have been well-studied in the psychology literature are increasingly found to play a role in how people understand data in visualizations~\cite{Bedek2017,Kong2018,Mansoor2017,Parsons2017,Valdez2018,Wall2017}. It is an important step to articulate how these biases manifest themselves not just at a broad, generalized level, but also with underrepresented populations, which may differ in their educational and socioeconomic profiles. 

Finally, there has been an emerging body of work that situates data visualization more specifically within the web's attention economy. Kim et al. found that visualizing personal and social predictions of data on the web can positively impact the recall and comprehension of that data~\cite{Kim2017, Kim2018}. Relevant to this work, a sequence of research by Kim et al. and Hullman et al. used simple analogies of physical measurements to aid in the understanding of abstract values~\cite{Kim2016, Hullman2018}. These studies suggest that presenting data in a personal, familiar manner may yield positive benefits to more diverse web audiences. 

\subsection{Why might rural populations be different?} 
Technological barriers and the ``digital divide'' remain as important, ongoing issues in rural America, where economic and infrastructure obstacles can fundamentally undermine the impact of emerging visualization tools and platforms~\cite{Burrell2018, Perrin2017, Rainie2004, Smith2013}. For example, research has found that financial factors in low-income communities often point to differences in access to devices - people with lower incomes tend to be more reliant on phones or public computers as their primary access points to the internet~\cite{Smith2013, Tsetsi2017}. Device constraints result in a host of problems for interactive data visualization, which often depend on screen real estate and mouse-based interaction~\cite{chittaro2006visualizing, ghose2012mobile, Ghosh, roberts2014visualization}. In addition, internet access may be intermittent, expensive, or unreliable, which can impact access to information resources. These challenges are mirrored in rural, central Pennsylvania (the focus of our study), where user access to high-speed internet can not be an assumption, and the long decline of the coal industry has brought with it gaps in income and education~\cite{Christofides2006, Glasmeier2008, Simeone2018, Yan2006}. 

Overcoming visualization literacy is a challenge in all contexts, but rural environments introduce additional constraints. Rural adult literacy programs face barriers due to financial resources and geographic isolation, resulting in less than 5 percent of the population being served by adult education programs~\cite{Yan2006}. As a result, large-scale education initiatives that necessitate widespread adoption and implementation may not be possible.

Together, these factors can drive inequality in the access of understandable data, and as a result, contribute to information inequality and ``the information poor''~\cite{Burrell2018, Yu2010, Yu2011}. Despite rural populations' under-representation in the data visualization literature, groups that mirror those in rural Pennsylvania compose of nearly 60 million people in the United States. It is with this in mind that we focus on understanding which factors drive attention towards (and away from) data visualizations.

\section{Study: Interviews in Rural PA}
We performed a series of interviews with community members in central Pennsylvania with the goal of capturing ~\textit{initial perceptions} of data visualization. 

\subsection{Stimuli}
To create a basis for discussion, we selected 10 different data visualizations that broadly involve the impact of drugs in the United States (Figure~\ref{fig:graphs}). These charts were chosen to represent a diverse set of features, including form, visual appeal, and source (Table~\ref{tab:graphs}). For the context of this study, we define \textit{source} as the location in which the chart was discovered, not  the origin of the data. Each chart was presented to participants in color on individual sheets of paper. 

\begin{figure}[h]
\includegraphics[width=\columnwidth]{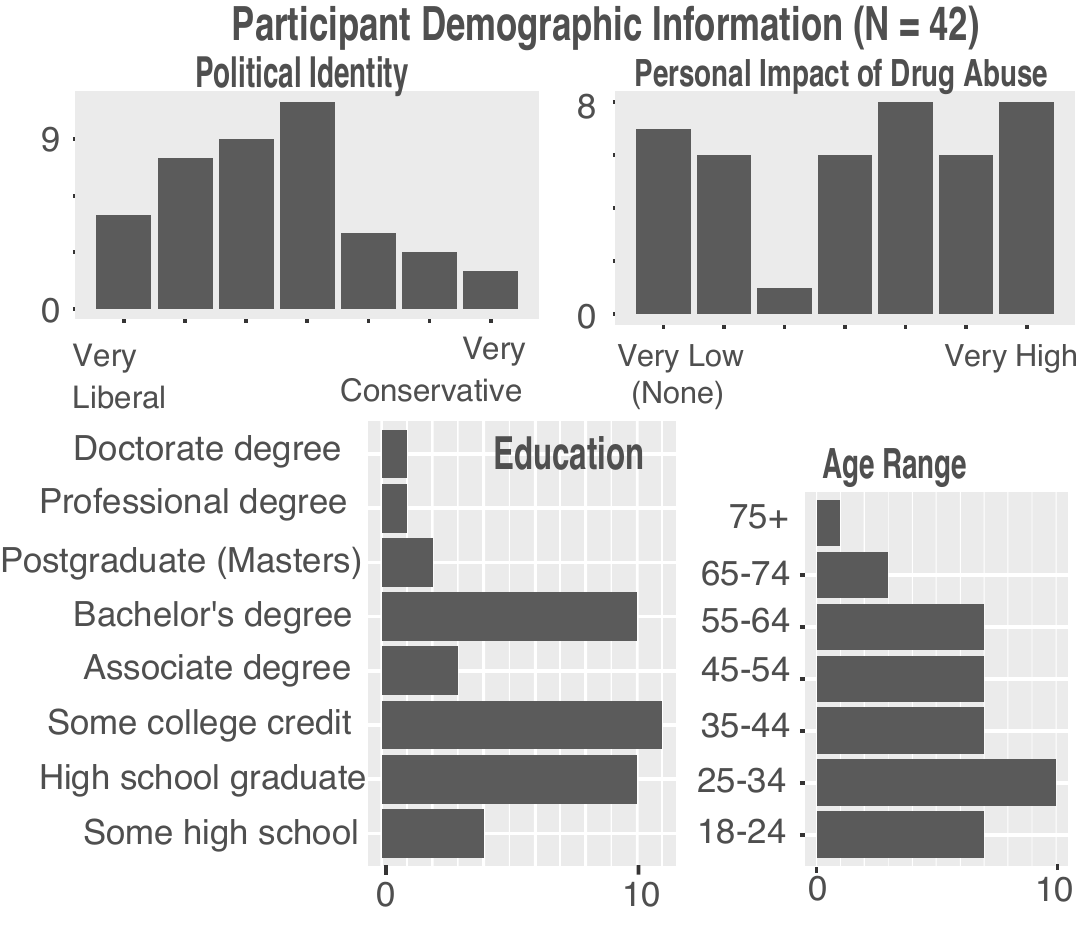}

\caption{Demographics of the community members we interviewed. Our participants represented diverse educational backgrounds and ages.}
\label{fig:demos}
\end{figure}

\subsection{Participants}
We interviewed 42 participants from three locations: 
\begin{itemize}
\item Staff members at a local university (13). Participants largely identified as working in food services as \textit{cashier}, \textit{line server}, \textit{prep kitchen}, or \textit{management}. 
\item Employees at a local construction site (5). Participants largely identified as working in \textit{demolition} or \textit{labor}. 
\item Visitors of a local farmers market (24). Participants were diverse in their backgrounds and occupations. 
\end{itemize}

For each participant, we collected their age, school district, political affiliation (``very liberal''(1) to ``very conservative''(7)), familiarity with graphs and charts, educational background, and the extent to which they had been personally impacted by drugs and/or addiction (see Figure~\ref{fig:demos}). Of the 28 participants that offered optional information about family income, 12 reported combined family incomes of less than \$45,000 a year. 7 reported family incomes of more than \$85,000 per year. 

\subsection{Procedure}
We followed a semi-structured interview process that varied depending on participant responses. Below, we give an overview of the interview structure, as well as the number of participants that were asked each question. 

\vspace{0.2cm}
\noindent
\textit{1. Introduction and consent.}

\begin{figure*}[h!]
  \begin{center}
  \includegraphics[width=\textwidth]{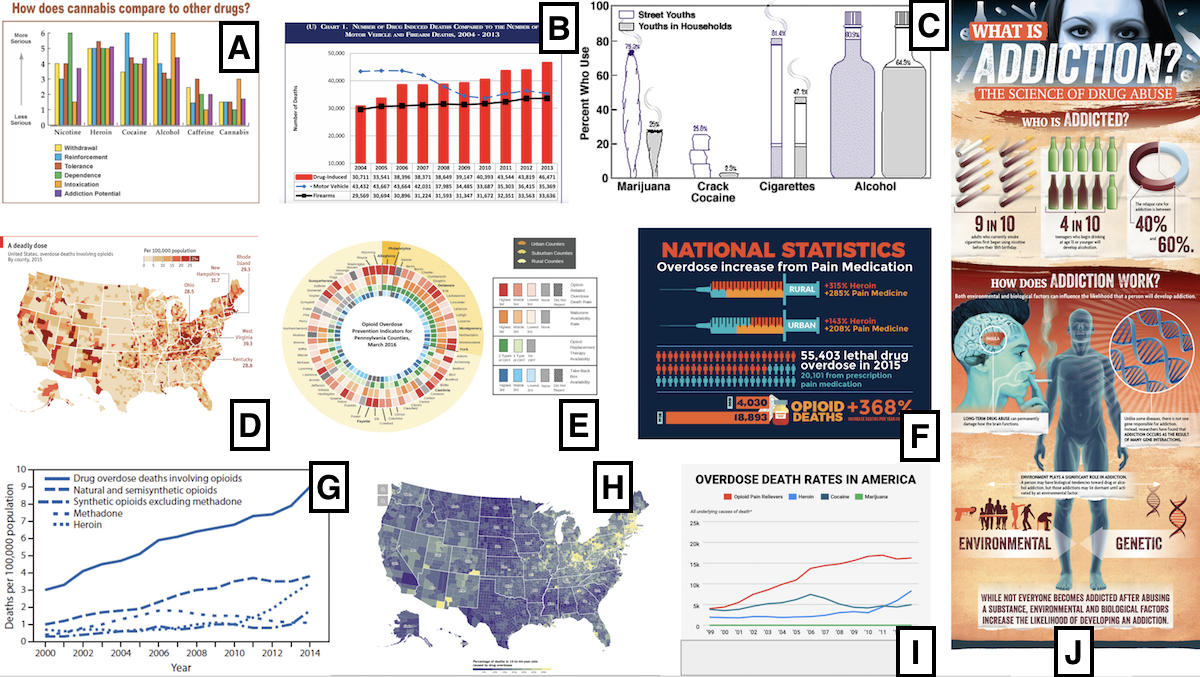}
  \caption{The graphs shown to participants. Each graph was presented on an independent sheet of paper}
  \label{fig:graphs}
  \end{center}
\end{figure*}

\noindent
\textit{2. Graphs presentation and ranking}. The 10 graphs in Figure~\ref{fig:graphs} were presented on 10 sheets of paper. Participants were given as much time as needed to consider them before responding to the following prompt: ``Based on \textbf{how useful they are to you}, arrange the graphs from \textit{most} useful to \textit{least} useful''. The framing of this prompt is critical and can guide participants in a number of directions. Based on a series of pilots, we found that `useful' was most successful in encouraging participants to express their own values in the context of data visualizations.

Depending on the rankings that the participants gave, we introduced a number of follow-up questions. The most common were prompted by similar graphs that were ranked differently:
\begin{itemize}
\item If line graphs were ranked differently, why? (N=35)
\item If maps were ranked differently, why? (N=22)
\item If infographics were ranked differently, why? (N=16) 
\end{itemize}
\noindent
\textit{3. Sources are revealed:} The sources of the graphs were revealed on the paper (Table~\ref{tab:graphs}). The researcher provided a brief background of unfamiliar sources (e.g. ``The National Vital Statistics System is part of the U.S. government's Centers for Disease Control and Prevention (CDC)''). After revealing the sources, participants were asked if they would like to change their rankings, as well as their rationale.

\vspace{0.2cm}
\noindent
\textit{4. Demographics questions:} Demographics were collected \textit{after} the interview to prevent priming participant responses.  

\vspace{0.1cm}
Interviews typically lasted approximately 15 minutes. Participants were given \$10 as compensation. This study was approved by the IRB at Bucknell University. 

\begin{figure}[b]
\begin{center}
\includegraphics[width=\columnwidth]{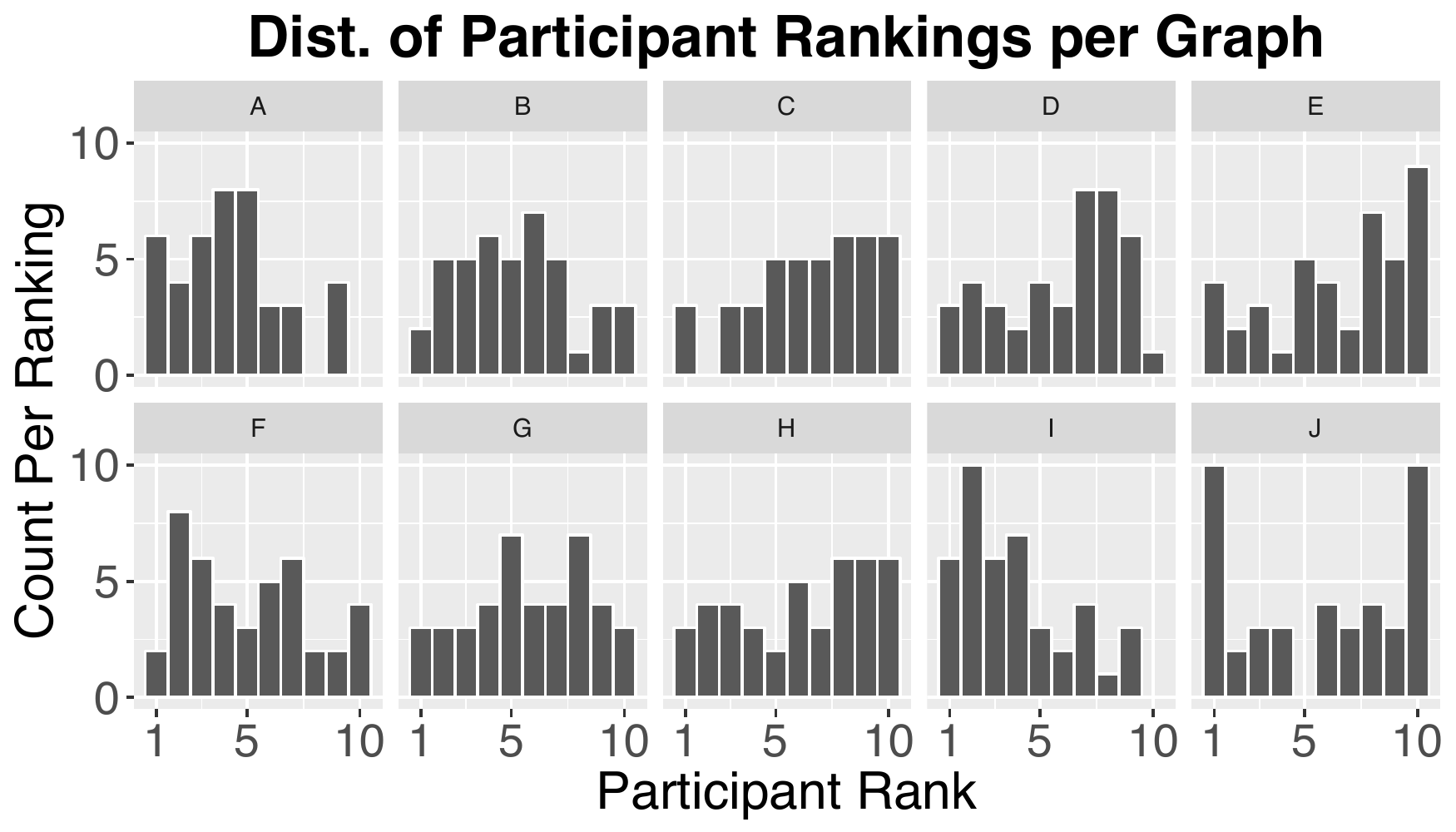}
\caption{Distribution of rankings given to each chart. While graphs provoked diverse opinions, infographics yielded the most polarized responses from participants (Graphs F and J)}
\label{fig:rank-dist}
\end{center}
\end{figure}

\begin{table*}[h!]
\begin{tabular}{lp{4cm}lp{4cm}p{6cm}}
\textbf{\#} & \textbf{Topic} & \textbf{Type} & \textbf{Found on (Source)} & \textbf{Perceptions (Code Frequency)}\\ \hline
A  & Severity of cannabis vs. other drugs      &  Bar    &  National Institute on Drug Abuse (NIDA)  & \textit{Relatable}(4), \textit{Informative}(2)      \\ \hline
B  & Comparison of drug, vehicle, and firearm deaths over time      & Bar / Line     &  BreitBart  & \textit{Confusing}(2), \textit{Informative}(2)      \\ \hline
C  & Drug use in `street' youths vs. youths in households      &  Isotype    &  National Institute on Drug Abuse (NIDA) & \textit{Simple}(3), \textit{Not trusted}(3), \textit{Clear}(2), \textit{Relatable}(2) \\ \hline
D & Overdose deaths involving opioids by county & Map & The Economist & \textit{Clear}(4), \textit{Attractive}(3), \textit{Confusing}(3), \textit{Cluttered}(3), \textit{Simple}(3), \textit{Relatable}(3)\\ \hline
E & Opioid overdose prevention indicators for PA counties & Heat map & Drexel University & \textit{Cluttered}(8), \textit{Confusing}(8), \textit{Clear}(4), \textit{Colorful}(4), \textit{Informative}(4) \\ \hline
F & Overdose increase from pain medication & Infographic & AgriMed (Medical Cannabis) & \textit{Attractive}(5), \textit{Confusing}(5), \textit{Simple}(4) \\ \hline
G & Drug overdoses over time & Line & National Vital Statistics System (NVSS) - CDC & \textit{Confusing}(6), \textit{Simple}(3), \textit{Cluttered}(2), \textit{Intriguing}(2) \\ \hline
H & Overdose deaths by country (15-to-44-year olds) & Map & The New York Times & \textit{Clear}(4), \textit{Colorful}(3), \textit{Relatable}(3), \textit{Simple}(3) \\ \hline
I & Overdose death rates over time & Line & Business Insider & \textit{Colorful}(16), \textit{Attractive}(6), \textit{Clear}(6), \textit{Simple}(5)  \\ \hline
J & The science of drug abuse & Infographic & Alternatives in Treatment (Rehab Center) & \textit{Informative}(4), \textit{Attractive}(3), \textit{Relatable}(3)
\end{tabular}
\caption{Graphs were chosen for representing diverse styles and sources. Codes are derived from interviews. When interpreting frequencies, recall that many participants chose to only comment on a select group of graphs}
\label{tab:graphs}
\end{table*}

\section{Analysis}
To associate responses with graphs, we use a process that is informed by the abbreviated version of grounded theory~\cite{charmaz2007grounded}. Instead of using existing visualization literature or theory to pre-generate a set of codes and themes, the research team generated  codes and themes by systematically going through the interview transcriptions. We use this particular process in order to mitigate the influence of assumptions from previous visualization research onto our group of participants. Our process was as follows: 

\begin{enumerate}
    \item Following the completion of interviews, transcripts were segmented to reflect the interview structure. Coding was performed section-by-section and further organized by graph.
    \item Two members of the research team iteratively analyzed the the transcripts, generating codes and themes until it was believed that a saturation point was reached. 
    \item Independently, two additional members of the research team used these codes to label all 42 transcripts. During this process, they generated new codes in order to best capture responses. 
    \item This data was returned to the two members of our research team described in (2), who revised the codes and themes based on this input. 
\end{enumerate}

The aggregated tallies of these codes, as well as the material of our study, can be found in our open repository: \url{https://osf.io/uxwts/}. When interpreting the number of instances of each code throughout the paper, it's important to remember that due to the semi-structured nature of our interviews, many participants did \textit{not} make direct comments about most graphs. 

\subsection{Rankings Overview: Clarity, Simplicity, and Color}
To give an overview of participant rankings after viewing the 10 data visualizations, Figure~\ref{fig:rank-dist} shows the distribution of \textit{initial} rankings of each graph. The most common codes associated with graphs across our interviews are as follows:  \textit{Colorful} (29)~\footnote{This value is likely inflated due to a question in which participants compared the two lines graphs - 16 of the mentions are about \textit{Graph I}}, \textit{Confusing} (29), \textit{Clear} (26), \textit{Simple} (26), \textit{Relatable} (21), \textit{Attractive} (20), \textit{Informative} (19), \textit{Cluttered} (17)

At this high level, we broadly see themes of simplicity, clarity, and aesthetics that align with other categorizations in data visualization~\cite{harrison2015infographic, Kennedy2017}. While tensions between highly stylized visualizations that emphasize engagement and minimal designs that prioritize understanding echo `chart junk' debates that have simmered for years in the HCI and visualization communities~\cite{Bateman2010, Borkin2013, Borkin2015}, we observed trends among our participants that suggest they gravitated towards straightforward visual encodings. Simple bar graphs (Graphs A, B) and line graphs (Graph I) emerged as among our more highly ranked charts. For example:

\vspace{0.2cm}
\noindent
\textit{``I think the ones with the pictures or the bars and fewer lines would probably make sense to more people, and the more details you get probably it get more convoluted to people''} -- P2, 35-44 year old, college degree
\vspace{0.2cm}

We resist sharing over-aggregated views of the data at this juncture for a simple reason - the distribution of rankings in Figure~\ref{fig:rank-dist} demonstrates diverse viewpoints that could be easily hidden within summary views. As a simple example, we found that perceptions of clarity typically did \textit{not} refer to a deep understanding of the data, but an ability to quickly extract the gist of the data, which often involved factors independent of the visual encoding. For P7, clarity was topical in nature, telegraphed by the visual presence of a brain representing addiction in Graph J:

\vspace{0.2cm}
\noindent
\textit{``...if you're just walking past, and you see it and you say `oh that has to do with the brain; I wanna stop and read that'. But you know like with this one [points to Graph G], you just say `oh yeah it goes up; it goes down''} - P7, 55-64 year old, high school graduate
\vspace{0.2cm}

Comparisons between the two line graphs - Graph G and Graph I - demonstrate how critiques of clarity and aesthetics often blurred together for our participants. 16 participants identified color as a distinguishing factor in their prioritization of Graph I over Graph G (Table~\ref{tab:graphs}), but they were often ambiguous as to whether color referenced general appeal or an improved visual encoding.

\vspace{0.2cm}

\noindent
\textit{``Color. The color makes it stand out more''} -- P17, 25-34 years old, high school graduate
\vspace{0.2cm}

While the terms `confusing' and `cluttered' were commonly used to critique data visualizations (29 and 17 times, respectively), many participants struggled to articulate any further reason why they disliked a particular chart. In the following sections, we will look into individual factors that help lend insight into these unspoken values and perceptions of data visualizations. 

\subsection{Data is Personal} 
Regardless of style, clarity, or ease-of-understanding, our interviews served as a reminder that data can be intimate and personal, and that those ties may supersede many other dimensions of design. We found more than 20 instances in which participants referenced a relatable component of the graph's content (the code \textit{relatability}). Consider the following participants discussing why they valued Graph J - our only chart that makes reference to alcohol: 

\vspace{0.2cm}
\noindent
\textit{``Information about alcohol right there [in Graph J]. I was a functioning alcoholic. The most important person in my life is an alcoholic. Right now that's important to me''} -- P14, 65-74 year old, college graduate

\vspace{0.1cm}
\noindent
\textit{``Well [Graph J] obviously gets me because I drink and smoke.''} -- P30, 25-34 year old, some high school but no diploma

\vspace{0.2cm}

Like much of America, the opioid crisis has been especially destructive in central Pennsylvania - a region that has experienced economic hardship in some sectors alongside the decline of the coal industry. Our participants were no exception, and the importance of communicating the dangers of opioids emerged repeatedly during our interviews. 

\vspace{0.2cm}
\noindent
\textit{``I have a few friends that died [from drug overdoses] so [Graph F] made me put it that way''} -- P29, 25-34 year old, high school graduate

\vspace{0.1cm}
\noindent
\textit{``I picked [Graph E] because this has different counties. And you can see which counties have the most opioids... I put [Graph G] because it's a line graph and once again it points out that opioid is the number one cause of death...basically what it comes down to is that opioids is the number one killer in Pennsylvania''} -- P34, 25-34 year old, high school graduate
\vspace{0.1cm}

\noindent
\textit{``As for some of the other [graphs], I happen to know quite a few people who unfortunately happen to have an issue with opioids... and it's something you consider... are you going to see that person tomorrow or not?''} -- P30, 25-34 year old with some high school, but no diploma
\vspace{0.2cm}

Personal experiences not only molded perceptions of graphs that contained opioid data, but also graphs that did \textit{not} reference opioids. P20 said the following about disliking Graph J:

\vspace{0.1cm}
\noindent
\textit{``Because [Graph J] doesn't show anything on it about opioids and I think that is one of the biggest problems. It shows marijuana, crack cocaine, cigarettes and alcohol. You don't see many people dying from that.''} -- P20, 18-24 year old with some high school, but no diploma
\vspace{0.1cm}

At the conclusion of the interview, participants were asked to rate on a 1-to-7 scale how they have been personally impacted by drug abuse (either themselves or people they care for). 22 of our 42 participants responded with a 5 or higher, including 8 participants responding with a 7 out of 7 (Figure~\ref{fig:demos}). For most of our participants, these experiences were unspoken and not captured by our interviews. And yet, for those that were impacted so significantly, it is possible that they trumped every other factor we analyze in this paper.

\subsection{Geographical Information: Where am I in the data?}
In our stimuli, we presented two maps (D and H) - both of which were identical in many factors~\footnote{It's important to note that these maps were designed to be interactive, so while we can comment on perceptions about the static version of these images, user comments may \textit{not} translate to interactive versions.}. Perceptions of maps were diverse, with some participants believing them to be clear (coded 8 times) or simple (6) representations of data and others believing them to be confusing (5) or cluttered (5).

\vspace{0.2cm}

\noindent
\textit{``The ones with the picture of the map, that's kind of a cluster. To me, I just want to pick it up and read it. I do not want to have to just looks and see, follow the arrows to the different colored areas''} - P16, 45-54 year old, college degree
\vspace{0.2cm}

However, the notion of the personal manifested itself again in perceptions of these graphs.  When participants commented about the geographical data, we found at 6 instances in which conversation focused on locations where they either live currently (looking for `home') or had lived previously: 

\vspace{0.2cm}
\noindent
\textit{``This one was interesting because I used to live in West Virginia ... and drugs were becoming very bad in that area ... so I guess with seeing the states and where problems are kind of caught my eye.''} - P25, 45-54 years old, high school graduate
\vspace{0.1cm}

\noindent
\textit{``It's just a little more congested; it's easier to read the others vs trying to pinpoint where in a state I would be''} - P6, 45-54 year old, Associate's degree
\vspace{0.1cm}

\noindent
\textit{``I think in this one, Pennsylvania stuck out more to me than this one.''} - P20, 18-24 year old, college student
\vspace{0.1cm}

\noindent
\textit{``These two [maps] are probably here because I like them less. It's the whole country; it's so huge. You naturally look at your state. It's too busy. I'm not that thrilled with those.''} -- P37, 65-74 year old, high school graduate
\vspace{0.1cm}

These perceptions held for graphs with any indications of geography. For example, although both line graphs were about drug abuse in the United States, Graph I was more clearly marked with the title ``Overdose Death Rates in America'', leading to this exchange:

\vspace{0.2cm}
\noindent
\textit{
``[I ranked Graph I higher] just for the simple fact that I live in America so I thought it was pretty relevant... more than the other one [Graph G].''} - P27, 25-34 year old, high school graduate
\vspace{0.2cm}

Finally, even for graphs which were largely disliked by participants, finding a reference to their personal context made a dramatic impact in their ranking. Consider P37's justification for giving Graph E's county heat map a high ranking (our lowest ranked graph among participants):

\vspace{0.1cm}
\noindent
\textit{``It has my own county in it. If this would have been like all the countries of the world or something, I really wouldn't have been interested in it because it just affects me personally. So I think because it is something that I can personally relate to.''} -- P37, 65-74 year old, high school graduate
\vspace{0.1cm}

While national trends with larger sample sizes may more accurately communicate drug use and abuse in the United States, our participants' focus on their local region may have design implications for rural populations (and perhaps more broadly).

\subsection{Social Framing: Will this help \textit{other} people?}
Despite our prompt, we found that a couple of participants oriented their comments in a more external, social framing. Instead of ranking charts and graphs based on which forms were most effective to them, these participants were concerned about the effectiveness of charts and graphs for \textit{other} people. For example, while P4 personally gravitated towards infographics, they worried that \textit{other} people would not:  

\vspace{0.2cm}

\noindent
\textit{``A lot of times, I think, infographics are taken less seriously than a `straight chart' because they're a newer way of visualizing. That's helpful in getting your attention, but that's what made them fall [in my rankings]''} -- P4, 25-34 year old, college degree
\vspace{0.2cm}

Similarly, P31 - a school principal - ranked infographics the highest because of a belief that the graphical nature would be \textit{more} effective for students and their parents:

\vspace{0.2cm}
\noindent
\textit{``I ranked them based on the fact that I am a [school] principal and [which graphs] I would be wanting to show to my kids and parents; It was based on the information that's provided, and then also the appeal; the visual aspect of it and what is going to engage them''} -- P31, 35-44 year old, postgraduate degree
\vspace{0.2cm}

While this framing differed from the task we provided, we should consider that the effect was amplified given the topic at hand. Many participants likely see the impact of opioid abuse directly in their home communities. As a result, the urgency of communicating that information locally may be heightened. 

\subsection{Infographics: Clarity and Novelty}
Infographics have historically been a divisive mode of data communication, and our interviews were no exception - Graph J received the most polarizing rankings of any chart (Figure~\ref{fig:rank-dist}). As the codes in Table~\ref{tab:graphs} reveal, participants who had positive feelings about infographics (Graphs F and J) found them to be clear (5), simple (5), and attractive (8), aligning with previous analyses of infographics~\cite{harrison2015infographic}. However, participants that were critical of infographics were more diverse in their rationales, often struggling to articulate why they placed it so low in their rankings. Those participants called them ``boring'', ``bizarre like a science fiction comic book'', ``childlike'', and ``not serious'' (among others). 

\vspace{0.2cm}
\noindent
\textit{``I just don't like that graph because it reminds me of something you would see in a magazine and not necessarily something you would see in a science article.''} -- P41, 18-24 year old, college graduate
\vspace{0.2cm}

\noindent
\textit{``It's like the guy had access to pictures and he didn't know when to stop''} -- P33, 45-54 year old, postgraduate degree
\vspace{0.2cm}

Additionally, P8 commented that being ``older'' made it harder to understand ``new'' information visualizations such as infographics: \textit{``My mind has to study it harder''}. While we did not observe this sentiment as a consistent trend in older participants, work by Harrison et al. found that infographics were often rated lower by older people~\cite{Harrison2013}. More broadly, these perceptions seem to align with observations about novices encountering data visualizations~\cite{Lee2016, Maltese2015, Grammel2010, Elias2011}, even as the infographics themselves may be designed for simplicity. 

\vspace{0.2cm}
\noindent
\textbf{Statistical familiarity:} To further investigate factors that may impact perceptions of infographics, participants who ranked F and J significantly differently were asked about the distinction they made. While some participants found Graph F to be simple (4), others found it confusing (5). One possible explanation may be related to statistical familiarity and comfort that comes from exposure through formal education. Consider the rationale given by P16: 

\vspace{0.2cm}
\noindent
\textit{``This one over here [Graph F] has a needle, but understanding 315\% or 285\% - that's not realistic to me. I'm used to 10-100 percent.''} - P16, 45-54 year old, college degree
\vspace{0.1cm}

 While P16 was a college-educated member of the community, the other three participants who commented on the confusing nature of Graph F were not. By contrast, all four participants who commented on the \textit{simplicity} of the same graph either held a college degree (3) or frequently interacted with numerical data in their work (an office manager).

\subsection{Why \textit{don't} people change their rankings?}
After the sources of each data visualization were revealed (listed in Table~\ref{tab:graphs}), participants were given an opportunity to change their rankings. However, 25 of 42 participants opted to keep their initial rankings. We found that the rationale of 22 participants could be described in four categories:

\begin{itemize}
\item \textit{Source is irrelevant (9):} expressed that the source does not impact the data and/or presentation. 
\item \textit{Ranked on other criteria (5):} expressed that their initial ranking was based on other criteria (visuals, interest) and that criteria had not changed.
\item \textit{No reason (4):} could not (or was not willing to) articulate any reason for maintaining their rankings
\item \textit{All sources are trusted (3):} perceived that all sources were equally trustworthy. 
\end{itemize}

\begin{figure}[h]
\begin{center}
  \includegraphics[width=\columnwidth]{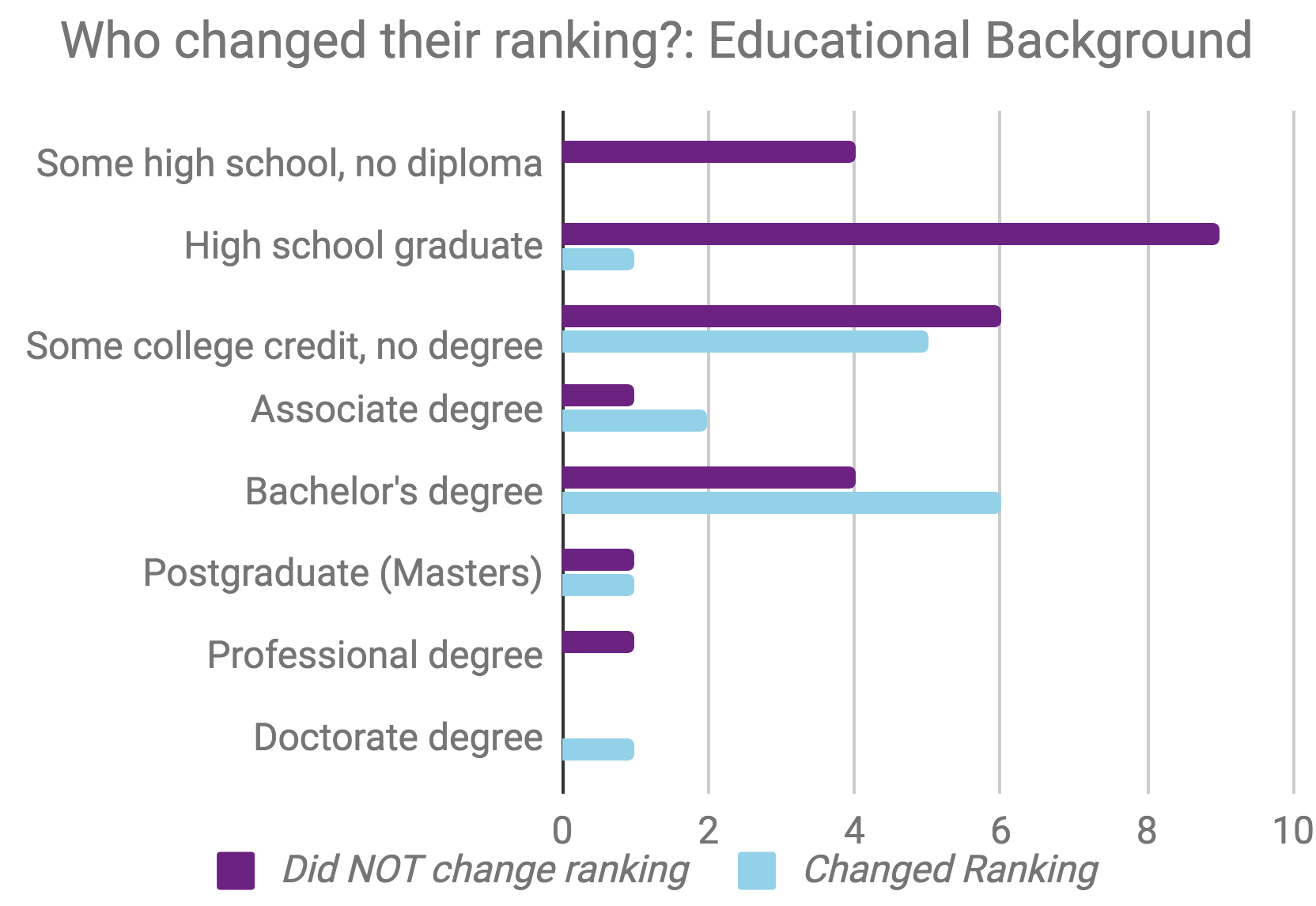}
  \caption{More educated participants were more likely to change their rankings after seeing the graph's source}
  \label{fig:ed-change}
 \end{center}
\end{figure}

\vspace{0.2cm}
\noindent
\textbf{Education:} The decision not to change rankings aligned the educational background of our participants (Fig.~\ref{fig:ed-change}). Of the 25 participants who maintained their rankings, 52\% (13 out of 25) never received post-secondary education (partial or completed). In contrast, only 1 out of the 17  participants who changed their rankings fell into the same category. While we hesitate to offer conclusions based on this sample, it reinforces observations that results from studies sampling educated participants may not generalize to other populations.  

\vspace{0.2cm}
\noindent
\textbf{Objectivity of data:} 12 of 22 participants that did not change their ranking either expressed beliefs that the source of the data visualization is irrelevant or that all sources were equal.

\vspace{0.1cm}
\noindent
\textit{``I think that information is information no matter from where it comes from.''} -- P20, 18-24 year old with some college credit, no degree

\vspace{0.1cm}
\noindent
\textit{``The origin doesn't really matter to me, it's just the information they are giving you. If it's not clear and not concise I don't care where it came from''}  - P14, 65-74 year old, college degree
\vspace{0.2cm}

The idea that ``information is information'' may suggest that people subscribe an objective characteristic not only to data, but also to visualization designs that communicate data. 

\vspace{0.2cm}
\noindent
\textbf{Reluctance to change:} Related to the observations above, 5 of 22 participants expressed attitudes suggesting that they would not change their rankings, regardless of what new information was introduced to them:

\vspace{0.2cm}
\noindent
\textit{``I did it based on how they were easiest to read and comprehend, so it doesn't really matter the source as far as when you are able to read it and comprehend it. Whether it's a `business' one or a `treatment' one, I don't think it really matters that much''} -- P2, 35-44 year old, college degree
\vspace{0.2cm}

Taken together, these perceptions suggest that the final rankings in this study may not reflect participant rankings had they been given source information in advance.  For example, when asked why he did not change his rankings, $P15$ revealed he \textit{``doesn't often change his mind''}. This attitude may reflect an \textit{anchoring effect}, in which the first information that people see can dramatically impact later answers. For some participants, this effect held even as people simultaneously expressed beliefs about the importance of the source of data. Consider $P8$, who opted \textbf{not} to change her ranking: 

\vspace{0.1cm}
\noindent
\textit{``With the social media, the way it is today, it's hard to trust a lot of information because you could put anything on anything; and just because it's there doesn't mean it's accurate or it's truthful''} - P8, 55-64 year old high school graduate. 
\vspace{0.2cm}

Anchoring and priming effects have only more recently been studied in the context of data visualization, focusing primarily on the perceptual level~\cite{Harrison2013, Valdez2018}. However, if these observations generalize, it may have design implications for the communication of source information. 

\subsection{Why \textit{do} people change their rankings?}
17 of 42 participants changed their rankings after the sources of the data visualizations were revealed. Not surprisingly, these participants placed a greater emphasis on the validity of sources in comparison to their counterparts: 

\vspace{0.2cm}
\noindent
\textit{``Just because it looks good doesn't mean that the information is right, and it is more important to have the right information''} - P18, 18-24 year old, some high school but no diploma

\vspace{0.1cm}
\noindent
\textit{``For me, I don't trust things if I don't know where it comes from.''} - P4, 25-34 year old, college graduate.
\vspace{0.2cm}

Figure~\ref{fig:changeDist} shows the distribution of ranking changes by graph. The graphs most commonly ranked higher - \textit{C: National Institute on Drug Abuse (NIDA)}, \textit{E: Drexel University}, and \textit{G: National Vital Statistics System (NVSS)} - all came from academic or government sources that were heavily criticized in the first set of rankings. For example, \textit{C: NIDA}'s depiction of bar graphs uses crude images such as alcohol bottles or cigarettes was perceived to be `childlike', `not serious', and `lacking credibility'. However, once the source was revealed, 8 out of 17 participants ranked them higher.

Similar to NIDA and NVSS, a chart from Drexel University \textit{E: Drexel} was initially perceived to be the most confusing of all of our charts, but received a positive ranking alteration from 8 participants after the source was revealed. In this case, it's unclear whether Drexel's standing as an academic institution or its position as a familiar entity (Drexel is located in Pennsylvania) was the driving force with participants. Locality of source played a significant factor for at least one of our participants in their assessment of \textit{The New York Times}:

\begin{figure}
\begin{center}
  \includegraphics[width=\columnwidth]{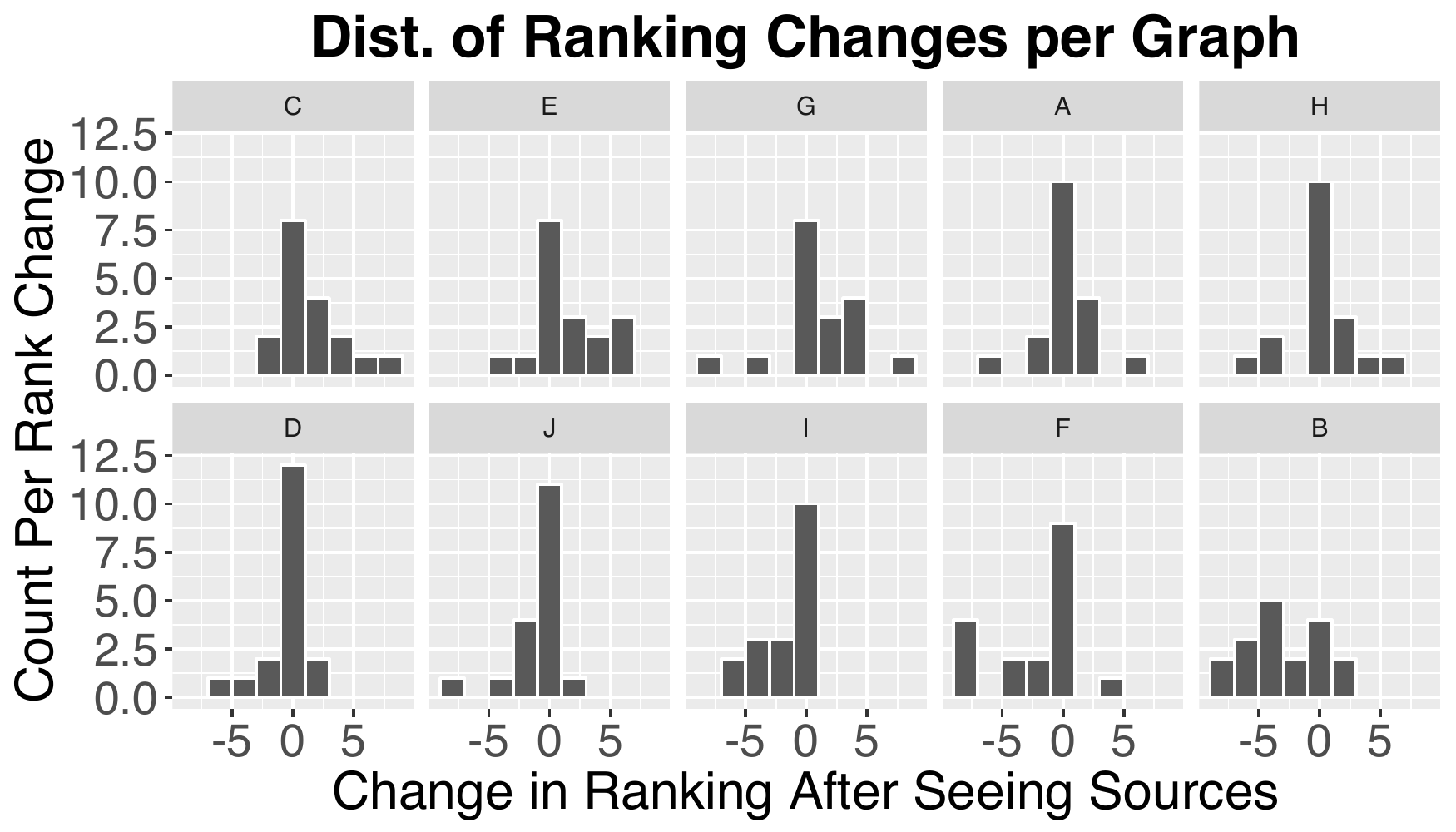}
  \caption{Distribution of how participants changed their rankings for each graph (N=17).  Positive shifts indicate \textit{improved} changes in rankings. Charts are ordered based on mean improvement in rankings.}
  \label{fig:changeDist}
 \end{center}
\end{figure}

\vspace{0.2cm}
\noindent
\textit{``I don't read this newspaper, but even if I did like this picture, I still won't buy the newspaper because I don't live in New York; The Sunbury paper, that's close to here. Then I would read it, but I still won't read that one''} -- P7, 45-55 year old, Associate's degree
\vspace{0.2cm}

P7's perspective alone is clearly not enough to make generalizing claims. However, given the value that participants placed on personal associations throughout our study, investigating the degree to which people prioritize visualizations from \textit{local} sources may offer valuable insight into how attention may be allocated towards (or away from) data.

\subsection{Trust and Political Identity}
While most participants valued governmental sources, notions of trust and distrust varied significantly between individuals. Consider the contrasting attitudes expressed by two participants' about the Center for Disease Control. 

\vspace{0.2cm}
\noindent
\textit{``I would never trust something from a drug company or Breitbart, and I would always trust something from the CDC or NIH''} -- P35, 75+ year old, Ph.D., identifies as very liberal. 
\vspace{0.2cm}

\noindent
\textit{``I don't trust the CDC. I believe they hide stuff.''} -- P16, 45-54 year old, Bachelor's degree, identifies as very conservative. 
\vspace{0.2cm}

\begin{figure}
\begin{center}
  \includegraphics[width=\columnwidth]{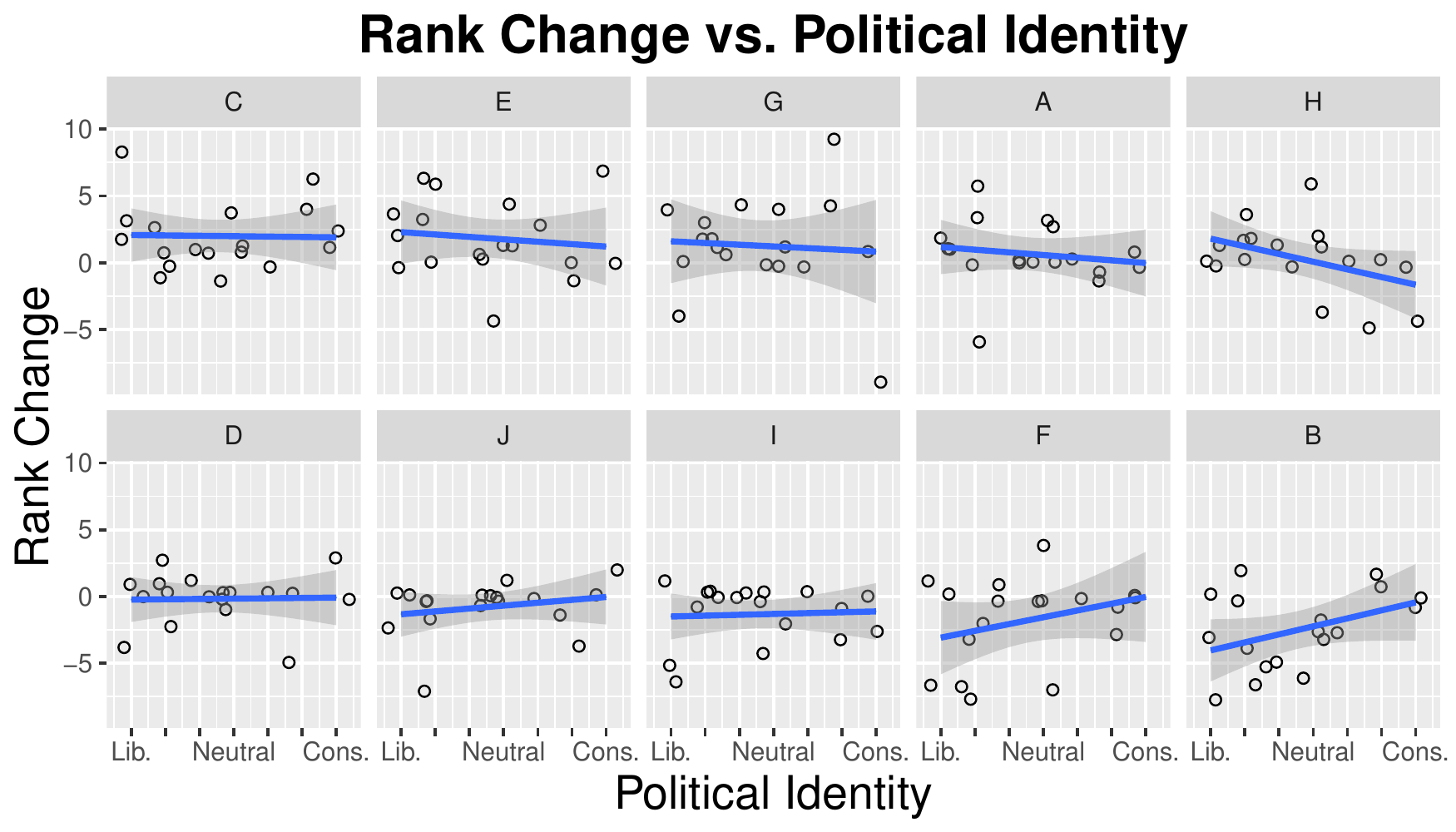}
  \caption{We found that rank changes in graphs tended to align with political identity in three of our graphs: H (The New York Times), F (AgriMed), B (Breitbart). Positive rank changes indicate \textit{improved} changes in rankings.}
  \label{fig:pol}
 \end{center}
\end{figure}

While these perceptions were not widespread among conservative participants, other graphs demonstrated a more partisan divide. \textit{H: The New York Times}, \textit{F: AGRiMED}, and \textit{B: Breitbart} all exemplified clearer trends that aligned with political identity (Figure~\ref{fig:pol}). Breitbart, for example, is a far-right news outlet that expectantly drew lower rankings from participants who identified as liberals. Surprisingly, we saw similar trends with AGRiMED, a licensed medical cannabis cultivation company - liberal participants were more likely to drop their ranking of the data visualization. Participants did not provide political affiliation information until \textit{after} completing the interview.

While these results are not surprising~\cite{arceneaux2012polarized, Pew2014}, we must consider how political biases and beliefs may negate the impact of even the most lauded institutions for data visualization design and implementation, such as \textit{The New York Times}. 

\section{Discussion}
In this study, we focused on how people in rural Pennsylvania allocate their attention towards or away from data visualizations - a decision-making process that precedes the sensemaking processes that are more often studied in the data visualization literature. Unfortunately, studies of this nature often leave us with more questions than answers. 

Our 42 interviews revealed a complex tapestry of motivations, preferences, and beliefs that impacted the way that participants prioritized data visualizations. We saw that participants valued clarity and simplicity, at times gravitating towards simple bar graphs and line charts. Other participants assigned high value to color and visual appeal. We found that highly educated people were more likely to value the source of a data visualization, and that trust can align with political identity. Each of these factors warrant more investigation, but here we will focus on the most dominant theme in our analysis - the impact of personal experience. 

\subsection{Visualizations are Personal}
The most recurring theme in our analysis were decisions framed or driven by personal experience. People that were impacted by abuse or addiction gravitated towards graphs that represented those substances. People often determined the quality of a graph containing geographical information based on how easy or difficult it was to find their home state. Other people tended to value graphs not based on their own understanding, but on the graph's perceived efficacy at communicating to \textit{other} people. Finally, one participant even suggested that they would pay more attention to visualizations from local news sources than national ones. 

Accounting for these factors will likely be critical in order to educate the general public through data visualizations, begging the question \textit{how can we design systems that align with the personal experiences of our audience?} There has been recent work in information visualization that paints with a broad stroke in this direction - from using more concrete analogies of distance~\cite{Kim2016, Hullman2018} to equipping interactive visualizations with search bars to enable the pursuit of personal goals~\cite{Feng2018}. 

The additional challenge posed in this research is that personal framings may fundamentally alter attention towards or away from a visualization \textit{before} those interactions occur. For people to engage with data, it may be necessary to not only provide mechanisms that allow people to explore personal dimensions of data, but to make sure those personal dimensions are front-and-center in first encounters with a visualization. 

The reluctance of people to change their rankings, coupled with perceptions of data objectivity (``information is information''), also suggest that first exposures to data may be critical. This presents a particularly challenging problem for designers as recent literature finds that people can be easily manipulated by a visualization's title alone~\cite{Kong2018}. However, given the prevalence of misinformation, these observations leave us with more important questions: \textit{How can platforms ensure that the first data visualization a person sees is a reliable one? How can we design data visualizations that are capable of altering previously formed impressions?}

Finally, the demographic indicators (in education and political identity) that aligned with participant priorities serve as a reminder that data visualization studies performed primarily with highly educated students may \textit{not} generalize to the larger public, echoing similar observations in psychology~\cite{henrich2010most}.

\subsection{Limitations}

Our observations should be understood within the design of our interviews - 42 participants reflecting on 10 graphs, many at a local farmers market. It is unclear how representative our sample is of hard-to-access, marginalized communities in the United States. While factors of age, family income, and educational backgrounds roughly reflected our expectations for the region, political affiliations skewed more liberal than voting records would suggest. 

The findings in this paper should be replicated with controlled studies, different populations, and larger sample sizes. Additionally, the framing of ``how useful are these visualizations?'' should be contrasted with other framings (``how effective?'' or ``how interesting?''). Further research in this direction will need to carefully target groups that are difficult to access through traditional means of recruitment. 

\subsection{Broader Impacts}

We have motivated this paper with the goal of communicating data to groups that are underrepresented in the visualization literature. However, research that articulates methods of communication to underrepresented groups also has the danger of providing a road map for manipulation through misinformation. We find this potential outcome deeply unsettling, but also one that is already steeped in reality through the infrastructure and incentive-structures of our online systems. For example, people may first see a visualization in social media or search results that are devoid of clear source information, or recommended based on our political affiliations. Our hope is that this research can be used to rethink the design of our information systems to combat misinformation rather than amplify it.

\begin{acks}
  The authors would like to thank Owais Gilani (Bucknell University), Aaron Quigley (University of St. Andrews), and Jennifer Silva (Bucknell University) for their guidance at various points of this project. 

\end{acks}

\bibliographystyle{ACM-Reference-Format}
\bibliography{chi_bib}

\end{document}